\documentclass[12pt,preprint]{aastex}

\shorttitle{Cool Gas and Massive Stars - the Nuclear Ring in M100}
\shortauthors{Allard et. al.}

\begin{document}


\title{Cool Gas and Massive Stars - the Nuclear Ring in M100}


\author{E. L. Allard}
\affil{Centre for Astrophysics Research, University of Hertfordshire, College Lane, Hatfield, Herts, \\AL10 9AB, UK}

\author{R. F. Peletier\altaffilmark{1}}
\affil{Kapteyn Astronomical Institute, University of Groningen,
	 Postbus 800, 9700 AV Groningen, The Netherlands}

\and

\author{J. H. Knapen}
\affil{Centre for Astrophysics Research, University of Hertfordshire, College Lane, Hatfield, Herts, \\AL10 9AB, UK}
\altaffiltext{1}{School of Physics and Astronomy, University of Nottingham, University Park, Nottingham, \\NG7 2RD, UK\\
Accepted for publication in ApJ Letters, 23rd September 2005}


\begin{abstract}
The SAURON integral field spectrograph was used to observe the central
area of the barred spiral galaxy M100 (NGC 4321). M100 contains a
nuclear ring of star formation, fueled by gas channeled inward by
the galaxy's bar. We present maps of emission line strengths,
absorption line strength indices, and the gas velocity dispersion
across the field. The H{$\beta$} emission is strongest in the ring,
along two curved bar dustlanes and at the ends of the bar. The Mg b absorption
line strength shows a younger population of stars within the ring as
compared to the surrounding area. The gas velocity dispersion is
notably smaller than elsewhere in the field both in the ring and along
the leading edge of the dustlanes. Low gas dispersion is correlated spatially with
the H{$\beta$} emission. We thus see stars being formed from cold (low
dispersion) gas which is being channeled inward along the dustlanes
under the influence of a large bar, and accumulated into a ring near
the location of the inner Lindblad resonances. This lends further
strong support to the interpretation of nuclear rings in barred
galaxies as resonance phenomena.

\end{abstract}



\section{Introduction}

The centers of many spiral galaxies show evidence of star formation
(SF), often organized into a nuclear ring or pseudo-ring. Such rings
are found in 20{$\% $} of spiral galaxies (Knapen 2005), are thus
relatively common, and almost always occur within a barred host (Buta
\& Combes 1996; Knapen 2005). They are sites of massive SF; Kennicutt
et al. (2005) estimate that nuclear rings contribute 3-5$\%$ to the
overall current SF rate in the local universe. Dynamically, nuclear
rings are thought to trace the position of the inner Lindblad
resonances (ILRs), where gas driven in under the influence of the
large bar slows down (Lynden-Bell \& Kalnajs 1972; Shlosman
1999). Because nuclear rings turn inflowing gas into stars within the
central region, they will contribute to secular evolution and pseudo-bulge
formation. The SF within the rings is most likely a combination of
gravitational instabilities (Elmegreen 1994) and the influence of
shocks and density waves (Knapen et al. 1995a, hereafter K95; Ryder,
Knapen \& Takamiya 2001).

 
M100 (NGC 4321) is a prominent, relatively face-on spiral galaxy with
a moderately strong bar, at a distance of 16.1~Mpc (Ferrarese et
al. 1996); at this distance 1 arcsec is equal to 70~pc. M100 hosts a
well-known nuclear ring with prominent massive SF, which is located
near a pair of ILRs induced by the bar (K95).  A detailed analysis of
the optical and NIR morphology of the circumnuclear region of M100 has
been given by K95, Knapen et al. (1995b) and Ryder \& Knapen (1999).


Here we present integral field spectroscopic data of the bar and
circumnuclear region of M100 and derive maps of emission line
intensities, gas velocity dispersions and absorption line
strengths. We report a ring of relatively cold gas within the
circumnuclear region, which lies exactly at the location of the
H{$\beta$} emission.


\section{Observations and Analysis}

The SAURON integral field spectrograph (Bacon et al. 2001) was used on
the 4.2m William Herschel Telescope (WHT) on La Palma on 2003, May
2. The field of view was 33$\times$41 arcsec$^{2}$, and contains 1431
square lenses of 0.94$\times$0.94 arcsec$^{2}$ in size, each producing
a spectrum. A further 146 lenses cover a region 1.9~arcmin from the
field for simultaneous observation of the sky background. The
wavelength range of 4760-5350{\AA} is observed, which contains the
stellar absorption features H{$\beta$}, Mg b, Fe5015, and the emission
lines H{$\beta$}, [O{\sc iii}] and [NI]. The velocity resolution is 110~km\,s$^{-1}$. To cover the complete bar and
circumnuclear region of M100, three pointings were made (see Figure 1f).

The data were reduced using the specially
developed XSauron software, originally described by Bacon et
al. (2001). The data were bias and dark subtracted and the spectra
extracted, wavelength and flux calibrated. The separate exposures were
merged and mosaicked together to produce one final datacube. This was
spatially binned using the Voronoi 2D binning method of Cappellari \&
Copin (2003) to achieve a minimum S/N across the field of 60 per angstrom for the
extraction of the kinematics, and 100 for the line strength maps.

To determine the gas kinematics and emission line strengths we followed a 2-step
approach (Sarzi et al. 2005): we first determined in each
bin the stellar line of sight velocity distribution (LOSVD) using the Penalized
Pixel Fitting Method (Cappellari \& Emsellem 2004), using the spectra outside
the region of the emission lines. We then fitted the full spectra using a linear
combination of single age, single metallicity models from Vazdekis (1999), a 6th
order polynomial to fit the continuum, and gaussians with variable widths,
centers and amplitudes, all convolved with the previously determined LOSVD. In
this way maps of the different absorption and emission lines, presented below,
were determined.


\section{Results}

\subsection{Intensity Maps}

Figures 1a and 1b show the intensity of the emission lines
H{$\beta$} and [O{\sc iii}] over the field of view. H{$\beta$}
emission mimics the well-known H$\alpha$ morphology (e.g., Knapen et
al. 1995b) and is most prominent in the nuclear ring (broken up into
hotspots), and within the H{\sc ii} regions in the western
field. H{$\beta$} emission can also be seen along the two main
dustlanes in the bar, as outlined in the \emph {B-R} image (Fig
1e). The [O{\sc iii}] emission peaks at the center of the galaxy, and
is weaker than H{$\beta$} elsewhere while still tracing the most
prominent H{\sc ii} regions.

\subsection{Line Indices}

The method explained in Sect.~2 has enabled us to determine the
absorption line indices of Fe~5015, Mg~b and H$\beta$ (Lick system;
Worthey 1994) with a systematic accuracy of about 0.1\,\AA\ (see Sarzi
et al. 2005 for detailed simulations to establish the accuracy of the
absorption - emission line separation).  The Mg b line absorption line
strength (Fig.~1c) increases for older and more metal rich stars. We
find a significantly lower value coinciding spatially with the
H$\beta$ emission, in both the nuclear ring and along the bar. From the [O{\sc iii}]/H$\beta$ values found for
this galaxy, the H$\beta$ emission in the ring is thought to come from
photoionization by stars, which have to be sufficiently hot, i.e. young. This
confirms that a young population of stars is dominates the spectrum
at these locations, diluting the stellar absorption features of the
older bulge stars. A more detailed analysis of the stellar populations
(including the Fe 5015 and H$\beta$ indices) within the central
regions of M100 will be presented by E. Allard et al. (in preparation, hereafter Paper II).

\subsection{Gas Velocity Dispersion}

The gas velocity dispersion map (Fig.~1d) shows a value of
$\sigma=140\pm$10 km\,s$^{-1}$ averaged over the central 5
arcsec. Barth, Ho \& Sargent (2002) measured a value of the stellar
velocity dispersion across a central 3.7~arcsec aperture of
$92\pm4$\,km\,s$^{-1}$. From our data, we find a value averaged over
the central 4~arcsec of $98\pm4$\,km\,s$^{-1}$ (Paper~II). These values are consistent with SAURON results in the literature (de Zeeuw et al. 2002; Emsellem et al. 2004; Sarzi et al. 2005) for a galaxy such as M100. They will include contributions from the rapidly rising rotation curve in gas (Knapen et al. 2000
measured a circular velocity of 145\,km\,s$^{-1}$ at a radius of
2~arcsec from the nucleus from H$\alpha$ Fabry-P\'erot data) and stars
(Paper~II, where we find a comparable value of some
140\,km\,s$^{-1}$), as well as from the unresolved broad line emission from the LINER nucleus. 

A ring of considerably lower velocity dispersion\footnote{Our quoted errors are derived from pixel to pixel variations within the field. Errors due to the determination of $\sigma$ within individual pixels are typically a few tens of km\,s$^{-1}$, depending on the value of $\sigma$ and the signal strength; see Sarzi et al. (2005) for a full discussion.},
$\sigma=51\pm$8 km\,s$^{-1}$, is seen at a distance of 7\,arcsec from
the center. It has a thickness of 7\,arcsec, and is surrounded by a
zone where $\sigma=102\pm$30 km\,s$^{-1}$ (Fig.~1d).  Regions of low
$\sigma$ are correlated spatially with enhanced
H{$\beta$} emission (Fig.~1d), such as the arc of material seen in the western side of the bar. An overlay of H{$\beta$} emission on
\emph {B-R} (Fig. 1e), shows that the low dispersion material
is {\it offset} by some 700\,pc from the dark dustlanes in the bar
(Sect.~4.1).

\section{Discussion}

\subsection{Gas Inflow}

Many barred galaxies show a symmetric pair of dustlanes, connecting
the nuclear region with the spiral arms, and associated with shocks in
the gas flow (Prendergast 1964; Athanassoula 1992).  At the location
of the shocks we expect strong shear, and SF will be
prevented. Immediately after the shock, however, the piling up of gas
allows material to cool and acquire a lower velocity dispersion, and SF
can be triggered. We see this expected offset, of some 700~pc in this
case, between the dustlanes and the ridges of SF, which coincides with
the cold gas (Fig.~1e). Zurita et al. (2004) studied the strong bar in
NGC~1530, and found that the largest velocity gradients occur at the
location of the dustlanes, with massive SF along the dustlanes offset
by some 900 pc.
    
\subsection{Creation of a Nuclear Ring}

Rings are thought to be the result of radial gas inflow (e.g., Schwarz
1979; K95; Heller \& Shlosman 1996). Material gathers in rings where
the radial flow of gas approaches an important resonance (set up by
the mass distribution of the bar). The nuclear ring emits strongly in
gas emission lines, confirming the existence of enhanced massive
SF. K95 modeled the gas flow in M100 using a self consistent disk of
gas and stars embedded within a halo. Using nonlinear orbital
analysis, a double ILR was confirmed to lie at the location of the
ring. Material collected at the ILR(s) experiences no net torque from
the bar pattern, and so the velocity dispersion of gas there will not
increase quickly with time, as may be the case outside the ring. The
fact that we see a circular region of lower dispersion, relatively
cold gas confirms this picture. Massive stars form in the ring which
implies that the gas in the ring is constantly being replenished.

\subsection{Star Formation in the Ring}

SF is generally thought to occur when the gas density of the ISM
exceeds the Toomre (1964) criterion for gravitational stability. Within
the ring, the density will continue to increase with the accumulation
of gas into the region. The ring will eventually become unstable to
collapse and SF is triggered (Elmegreen 1994), probably orchestrated
by the dynamics of the ring and bar (K95; Knapen et al. 1995b; Ryder
et al. 2001). The gas density within the ring of cold gas in M100 will
be greatest at the contact points between the ring and the dustlanes,
where gas flows into the region. We expect the distribution of the SF
and the youngest stars to be clumpy, and the intensity to be strongest
at these contact points. This is in fact the case, as can be seen in
Fig.~1b (see also in Knapen et al. 1995b). H{$\beta$} emission is an
indicator of very young stars ($\sim$10$^7$~yrs), and Mg~b of an older
population. The absorption line index of Mg b shows a complete ring, as
opposed to clumpy hotspots. This is probably because the
stars born from the cold gas diffuse slowly out to fill the ring (K95). The contact points are found at positions (-2,7) and (-2,-9) in Fig. 1c (offset RA and dec, in arcsec). The average Mg b index across both contact points (over apertures of 3 arcsec radius) is 0.98$\pm$0.18. We measure an average Mg b index across the rest of the ring of 1.34$\pm$0.1. These numbers confirm that indeed, as Fig. 1c suggests, the contact points have slightly lower Mg b than the rest of the ring (a 1.8 $\sigma$ effect), which suggests a younger
population at these positions.
 This supports the findings of Ryder et al. (2001) who found marginal evidence for an
azimuthal age gradient around the ring in M100, with the youngest star
forming regions found at the contact point. SF in the nuclear ring is,
in conclusion, likely to sequentially triggered, with the first burst
of SF triggered at the contact points, where the bar dust lanes
connect to the ring.

\section{Conclusions}

We have obtained integral field spectroscopic data of the central
region of the barred spiral galaxy M100. From these data we have
derived maps of emission line intensities, line strength indices, and
the gas velocity dispersion across the field. The H{$\beta$} emission
shows the well-known bright ring, broken into hotspots of
emission. The [O{\sc iii}] emission is brightest at the very center of
the galaxy, enhanced by LINER emission. The ring is seen as a clear
drop in the Mg b index values, suggestive of the dilution of the
stellar absorption features of the older bulge stars by the presence
of a younger population of stars within the ring. In the gas velocity
dispersion map, the ring is clearly colder than the underlying
disk. Significantly lower dispersion occurs exactly where the
strongest H{$\beta$} emission occurs, both within the ring and aligned
and offset from the dustlanes in the bar. The low gas dispersion
suggests that cold gas flows into the area through the dustlanes,
and accumulates into a ring under the influence of the
ILRs. Instabilities within this gas then trigger significant massive
SF. A more detailed analysis of the kinematics and stellar populations
within the region will be presented in Paper II.

\acknowledgments

We thank Martin Bureau and Marc Sarzi for insightful discussions, and
Isaac Shlosman and Tim de Zeeuw for comments on an earlier version of the manuscript.
Based on observations obtained at the WHT, operated on the island of
La Palma by the Isaac Newton Group in the Spanish Observatorio del
Roque de los Muchachos of the Instituto de Astrof\'{i}sica de
Canarias. We thank the SAURON team for making the instrument and the associated software available for this collaborative project. SAURON was made possible through
grants from the Netherlands Organization for Scientific Research, the
Institut National des Sciences de l'Universe, the Universities of
Lyon~I, Durham, Leiden, and Oxford, the British Council, the
Particle Physics and Astronomy Research Council, and the Netherlands
Research School for Astronomy. The real color image of M100 was kindly provided by Nik Szymanek. EA is supported by a PPARC studentship.



\clearpage

\begin{figure*}
\plotone{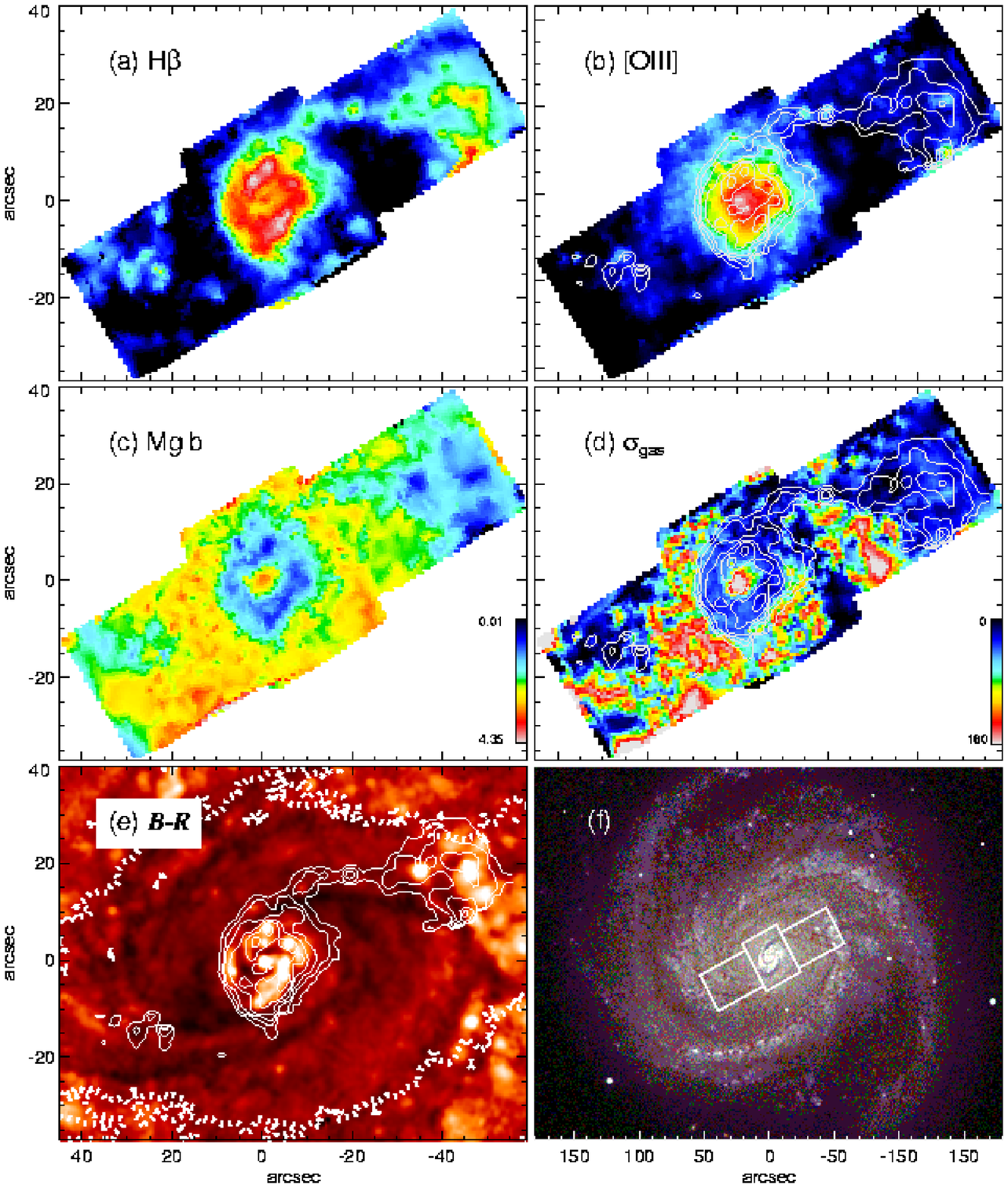}
\caption{SAURON maps of the central region of M100. (a). H{$\beta$} emission
line intensity. (b). [O{\sc iii}] emission line intensity; overlaid
are H{$\beta$} emission line contours at relative levels [0, 0.05,
0.1, 0.2, 0.5, 1, 2, 3]. (c). Mg~b absorption line strength. (d) Gas
velocity dispersion (in km\,s$^{-1}$); H{$\beta$} emission line
contours as in (b). (e). \emph {B-R} image from Knapen et al. (2004); darker shades indicate redder colors,
the location of the bar is indicated by a $K_{\rm s}$ band contour
(thick dashed line) at 18.3\,mag\,arcsec$^{-2}$; H{$\beta$} emission
line contours as in (b).  (f). Real color optical image of M100 with
approximate locations of the SAURON pointings, the size of this image is approximately $3.5 \times 4.5$ arcmin.  North is up and East is to the left.}
\end{figure*}

\end{document}